\documentclass[twocolumn]{revtex4}
\usepackage{graphicx}

\newcommand{\be}{\begin{equation}}
\newcommand{\ee}{\end{equation}}
\newcommand{\go}{\agt}
\newcommand{\lo}{\alt}
\newcommand{\bk}{{\bf k}}
\newcommand{\bB}{{\bf B}}
\newcommand{\zero}{{(0)}}

\begin{document}
\title{Polarized X-Ray Emission from Magnetized Neutron Stars:
Signature of Strong-Field Vacuum Polarization}

\author{Dong Lai}
\author{Wynn C.G. Ho}
\affiliation{Center for Radiophysics and Space Research,
Department of Astronomy, Cornell University,
Ithaca, NY 14853}

\begin{abstract}
In the atmospheric plasma of a strongly magnetized neutron star,
vacuum polarization can induce a MSW-like 
resonance across which a X-ray photon 
may (depending on its energy) convert from one mode into the other, with
significant changes in opacities and polarizations.
We show that this vacuum resonance effect
gives rise to an unique energy-dependent polarization signature in the surface 
emission from neutron stars: for ``normal'' field
strengths ($10^{12}\lo B\lo 7\times 10^{13}$~G), the plane of linear
polarization at photon energy $E\lo 1$~keV is perpendicular to that at $E\go
4$~keV, while for ``superstrong'' field strengths ($B\go 7\times 10^{13}$~G),
the polarization planes at different energies coincide.
The detection of polarized X-rays from neutron stars
can provide a direct probe of strong-field quantum electrodynamics 
and constrain the neutron star magnetic field and geometry. 
\end{abstract}

\maketitle

Surface emission from isolated neutron stars (radio pulsars and 
radio quiet neutron stars, including magnetars \cite{thompsonduncan95})
provides an useful probe of the neutron star (NS) 
interior physics, surface magnetic fields, and composition. The advent of 
X-ray telescopes 
in recent years has made
the detection and detailed study of NS surface emission a 
reality\cite{pavlovetal02}.
On the other hand, X-ray emission from accreting X-ray pulsars 
has long yielded information on the dynamics of accretion and radiative
mechanisms in the strong gravity, strong magnetic field
regime \cite{meszaros92}.

In the magnetized plasma that characterizes NS
atmospheres, X-ray photons propagate in two normal modes:
the ordinary mode (O-mode) is mostly polarized parallel to the $\bk$-$\bB$
plane, while the extraordinary mode (X-mode) is mostly polarized perpendicular
to the $\bk$-$\bB$ plane, where $\bk$ is the photon wave vector
and $\bB$ is the external magnetic 
field \cite{meszaros92}. This description of normal modes applies under typical
conditions, when the photon energy $E$ is much less than the electron 
cyclotron energy $E_{Be}=\hbar eB/(m_ec)=11.6\,B_{12}$~keV 
[where $B_{12}=B/(10^{12}~{\rm G})$], $E$ is not too
close to the ion cyclotron energy $E_{Bi}=6.3\,B_{12}(Z/A)$~eV (where $Z$ 
and $A$ are the charge number and mass number of the ion), 
the plasma density is not too close to the vacuum resonance (see below)
and $\theta_{kB}$ (the angle between $\bk$ and $\bB$) is not close to zero. 
Under these conditions, the X-mode opacity (due to scattering and
absorption) is greatly suppressed compared to the O-mode opacity,
$\kappa_X\sim (E/E_{Be})^2\kappa_O$ 
\cite{lodenquaietal74,meszaros92}.
As a result, the X-mode photons escape from 
deeper, hotter layers of the NS atmosphere than the O-mode
photons, and the emergent radiation is linearly polarized to a high degree 
(as high as $100\%$)
\cite{gnedinsunyaev74,meszarosetal88,pavlovzavlin00}.
Measurements of X-ray polarization, particularly when phase-resolved and
measured in different energy bands, could provide unique
constraints on the magnetic field strength and geometry and the compactness
of the NS \cite{meszarosetal88,pavlovzavlin00,heyletal03}.

\begin{figure}
\includegraphics[height=8cm]{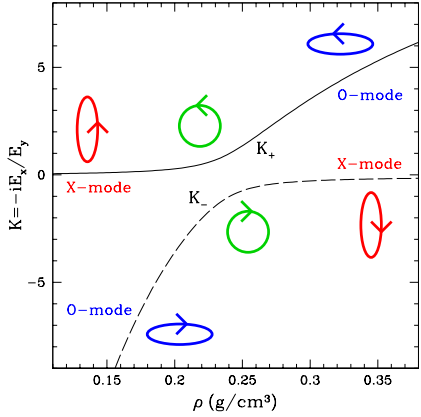}
\vskip -1cm
\caption{
The polarization ellipticity of the photon mode  
as a function of density near the vacuum resonance. 
The two curves correspond to the two different modes.
In this example, the parameters are $B=10^{13}$G, 
$E=5\,$keV, $Y_e=1$, and $\theta_{kB}=45^\circ$. 
The ellipticity of a mode is specified by the ratio $K=-iE_x/E_y$, 
where $E_x$ ($E_y$) is the photon's electric field component 
along (perpendicular to) the $\bk$-$\bB$ plane. The O-mode
is characterized by $|K|\gg 1$, and the X-mode $|K|\ll 1$.
\label{fig:f1}}\end{figure}

It has long been predicted from quantum electrodynamics (QED) that 
in a strong magnetic field the vacuum becomes birefringent
\cite{heisenbergeuler36,
schwinger51,adler71,tsaierber75,heylhernquist97}. Acting by itself, the
birefringence from vacuum polarization is significant (with the index of
refraction differing from unity by more than $10\%$) only for $B\go 300 B_Q$,
where $B_Q=m_e^2c^3/(e\hbar)=4.414\times 10^{13}$~G is the critical QED field
strength. However, when combined with the 
birefringence due to the magnetized plasma, vacuum polarization can greatly
affect radiative transfer at much smaller field strengths. A ``vacuum
resonance'' arises when the contributions from the plasma and vacuum
polarization to the dielectric tensor ``compensate'' each other
\cite{gnedinetal78,meszarosventura79,
pavlovgnedin84,laiho02}.
For a photon of energy $E$, the vacuum resonance occurs at the density 
$\rho_V\simeq 9.64\times 10^{-5}\,Y_e^{-1}B_{12}^2E_1^2 f^{-2}~{\rm g~cm}^{-3}
$, where $Y_e$ is the electron fraction, $E_1=E/(1~{\rm keV})$,
and $f=f(B)$ is a slowly varying function of $B$ and is of order unity
($f=1$ for $B\ll B_Q$ and $f\rightarrow (B/5B_Q)^{1/2}$ for $B\gg B_Q$; see
refs.\,\cite{laiho02,holai03}).
For $\rho>\rho_V$ (where the plasma effect dominates the dielectric
tensor) and $\rho<\rho_V$ (where vacuum polarization dominates), the 
photon modes (for $E\ll E_{Be}$, $E\neq E_{Bi}$ and $\theta_{kB}\neq 0$) 
are almost linearly polarized; near $\rho=\rho_V$,
however, the normal modes become circularly polarized as a result of the
``cancellation'' of the plasma and vacuum effects --- both effects tend to 
make the mode linearly polarized, but in mutually orthogonal directions.
When a photon propagates in an inhomogeneous medium, its
polarization state will evolve adiabatically (i.e. following
the $K_+$ or $K_-$ curve in Fig.~\ref{fig:f1}) if the density variation is
sufficiently gentle. Thus, a X-mode (O-mode) photon will be converted
into the O-mode (X-mode) as it traverses the vacuum resonance, with its
polarization ellipse rotated by $90^\circ$ (Fig.~\ref{fig:f1}).
This resonant mode conversion is analogous to the 
Mikheyev-Smirnov-Wolfenstein neutrino oscillation that takes place in
the Sun\cite{haxton95,bahcall03}. For this conversion to be effective, 
the adiabatic condition must be satisfied \cite{laiho02,laiho03}
(cf.~\cite{pavlovgnedin84})
\be
E\go E_{\rm ad}=2.52\,\bigl(f\,\tan\theta_{kB} |1-u_i|\bigr)^{2/3}
\!\left({1\,{\rm cm}\over H_\rho}\right)^{1/3}\!{\rm keV},
\label{condition}\ee
where $u_i=(E_{Bi}/E)^2$ and $H_\rho=|dz/d\ln\rho|$ is the density scale
height (evaluated at $\rho=\rho_V$) along the ray.
For an ionized Hydrogen atmosphere, $H_\rho\simeq 2kT/(m_pg\cos\theta)
=1.65\,T_6/(g_{14}\cos\theta)$~cm, where $T=10^6\,T_6$~K is the temperature,
$g=10^{14}g_{14}$~cm~s$^{-2}$ is the gravitational acceleration, and $\theta$
is the angle between the ray and the surface normal. In general, 
the mode conversion probability is given by \cite{laiho02,laiho03} 
\be
P_{\rm con}=1-\exp\left[-(\pi/2)(E/E_{\rm ad})^3\right].
\ee
The probability for a nonadiabatic ``jump'' is $(1-P_{\rm con})$. 

Because the two photon modes have vastly different opacities,
the vacuum resonance can significantly affect
the transfer of photons in NS atmospheres. 
When the vacuum polarization effect 
is neglected, the decoupling densities of the O-mode and X-mode photons 
(i.e., the densities of their respective photospheres, where the optical
depth measured from outside is $2/3$)
are approximately given by (for Hydrogen plasma and $\theta_{kB}$ not too 
close to $0$)
$\rho_O\simeq 0.42\,T_6^{-1/4}E_1^{3/2}G^{-1/2}$~g~cm$^{-3}$ and
$\rho_X\simeq 486\,T_6^{-1/4}E_1^{1/2}B_{14}G^{-1/2}$~g~cm$^{-3}$,
where $G=1-e^{-E/kT}$ \cite{laiho02}.
There are two different magnetic field regimes:
For ``normal'' magnetic fields,
\be
B< B_l\simeq 6.6\times 10^{13}\,T_6^{-1/8}E_1^{-1/4}G^{-1/4}~{\rm G},
\ee
the vacuum resonance lies outside both photospheres ($\rho_V<\rho_O,\rho_X$);
for the magnetar field regime, $B>B_l$, the vacuum resonance lies between
these two photospheres, i.e., $\rho_O<\rho_V<\rho_X$
(the condition $\rho_V<\rho_X$ is satisfied for all field strengths 
and relevant energies and temperatures). These two field regimes
yield qualitatively different X-ray polarization signals.

\begin{figure}
\includegraphics[height=8cm]{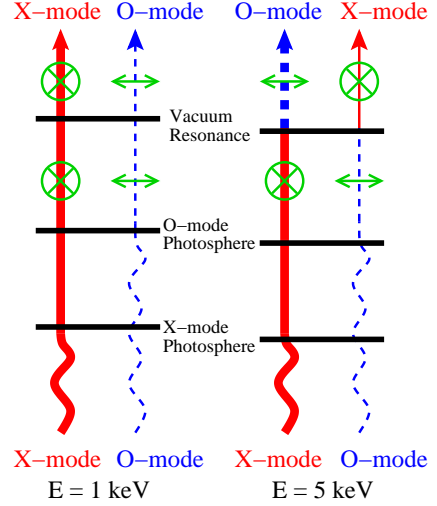}
\vskip -0.7cm
\caption{
A schematic diagram illustrating how vacuum polarization affects 
the polarization state of the emergent radiation from a magnetized 
NS atmosphere. This diagram applies to the ``normal''
field regime [$B\lo 7\times 10^{13}$~G, see eq.~(3)] in which the vacuum 
resonance lies outside the photospheres of the two photon modes.
The photosphere is defined where the optical depth (measured from outside)
is $2/3$ and is where the photon decouples from the matter.
At low energies (such as $E\lo 1$~keV), the photon evolves nonadiabatically
across the vacuum resonance (for $\theta_{kB}$ not too close to $0$), and
thus the emergent radiation is dominated by the X-mode. At high energies
($E\go 4$~keV), the photon evolves adiabatically, with its plane of
polarization rotating by 90$^\circ$ across the vacuum resonance, and thus
the emergent radiation is dominated by the O-mode. The plane of linear
polarization at low energies is therefore perpendicular to that at high 
energies.
\label{fig:f2}}\end{figure}

\begin{figure}
\includegraphics[height=7cm]{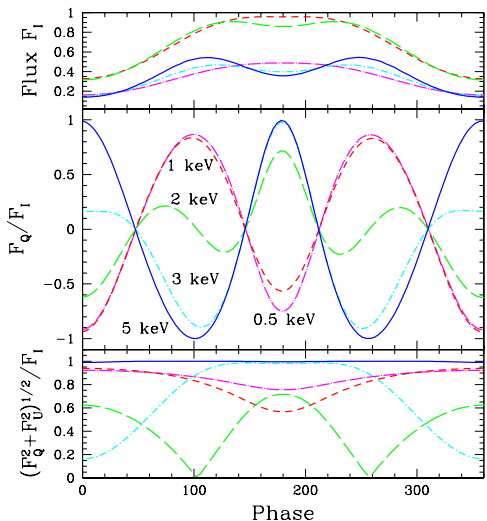}
\caption{
The phase evolution of the observed spectral flux 
(upper panel, in arbitrary units), the linear polarization $F_Q/F_I$ (middle
panel) and the polarization degree $P_L=(F_Q^2+P_U^2)^{1/2}/F_I$ (lower panel)
produced by the magnetic polar cap of a rotating NS. The model 
parameters are: the magnetic field $B=10^{13}$~G, the effective 
temperature of the polar cap $T_{\rm eff}=5\times 10^6$~K, the angle 
between the spin axis and the line-of-sight $\gamma=30^\circ$ and 
the angle between the magnetic axis and spin axis $\beta=70^\circ$.
The different curves are for different photon energies as labeled.
In the upper panel, the flux at 5~keV has been multiplied by 10 relative to 
the other curves.
The definition of the Stokes parameter is such that $F_Q/F_I=1$ corresponds
to linear polarization in the plane spanned by the line-of-sight vector and
the star's spin axis (the zero phase corresponds to the polar cap in the
same plane). Note that the polarization plane of high-energy 
($E\go 4$~keV) photons is perpendicular to that of low-energy photons
($E\lo 1$~keV). Also note that for $E$ around $E_{\rm ad}$, the polarization
degree $P_L$ can be zero at certain phases such that $P_{\rm con}=1/2$.
\label{fig:f3}}\end{figure}

\begin{figure}
\includegraphics[height=7cm]{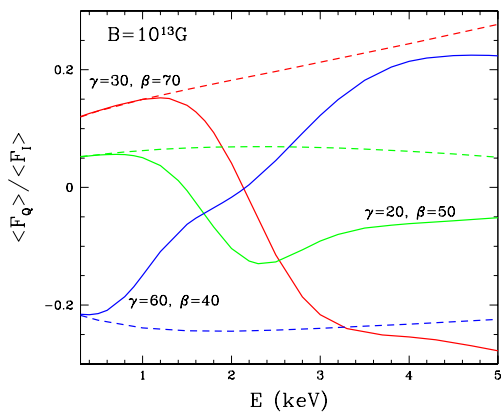}
\vskip -1.2cm
\caption{
The phase-averaged linear polarization as a function of
photon energy. The model parameters are $B=10^{13}$~G and
$T_{\rm eff}=5\times 10^6$~K, and the three curves correspond to
three different sets of angles ($\gamma, \beta$) as labeled. The averaged
$\langle F_U\rangle=0$. The dashed
curves depict the results when the vacuum polarization effect is 
turned off.
\label{fig:f4}}
\end{figure}

Consider the ``normal'' field strengths, $10^{12}\,{\rm G}
\lo B\lo B_l$, which apply
to most NSs (see Fig.~\ref{fig:f2}).
In this regime, the atmosphere structure
and total spectrum can be calculated without including vacuum polarization
[to be more accurate, we require $(B/B_l)^4\ll 1$ for this to be valid].
For concreteness, we consider emission from a hot spot (magnetic polar cap)
on the NS; the magnetic field at the hot spot (with size much 
smaller than the stellar radius) is perpendicular 
to the stellar surface. 
Let the specific intensities of the O-mode and X-mode emerging from their
respective photospheres (which lie below the vacuum resonance) be
$I_O^\zero$ and $I_X^\zero$, which we calculate using our H atmosphere models
developed previously \cite{holai01,holai03}. 
For a given $B$ and $T_{\rm eff}$, both $I_O^\zero$ 
and $I_X^\zero$ depend on $E$ and $\theta_{kB}$ at emission (the hot spot). 
As the the radiation crosses
the vacuum resonance, the intensities of the O-mode and X-mode
become $I_O=(1-P_{\rm con})I_O^\zero+P_{\rm con}I_X^\zero$
and $I_X=(1-P_{\rm con})I_X^\zero+P_{\rm con}I_O^\zero$. 
In calculating $P_{\rm con}$, we use the temperature profile of the atmosphere
model to determine the density scale height at the vacuum resonance.
We note that in principle, circular polarization can be produced when a photon
crosses the vacuum resonance \cite{laiho03}, but the net 
circular polarization is expected to be zero when photons from a finite-sized
polar cap are taken into account.

To determine the observed polarization, we must consider propagation of
polarized radiation in the NS magnetosphere, whose 
dielectric property in the X-ray band is dominated by vacuum polarization
\cite{heyletal03}. As a photon (of a given mode) propagates 
from the polar cap through the magnetosphere, its polarization state evolves
adiabatically following the varying magnetic field it experiences,
up to the ``polarization-limiting radius'' $r_{\rm pl}$, 
beyond which the polarization state is frozen. We set up a fixed coordinate
system $XYZ$, where the $Z$-axis is along the line-of-sight and the $X$-axis
lies in the plane spanned by the $Z$-axis and ${\bf\Omega}$ (the spin angular
velocity vector). The direction of the magnetic field 
${\bf B}$ following the photon trajectory 
in this coordinate system is specified by the polar angle
$\theta_B(s)$ and the azimuthal angle $\phi_B(s)$ (where $s$ is 
the affine parameter along the ray). The $X,Y$-components of the
electric field of the photon mode are $(E_X,E_Y)=(\cos\phi_B,\sin\phi_B)$
for the O-mode and $(-\sin\phi_B,\cos\phi_B)$ for the X-mode. 
The adiabatic condition
requires $|n_X-n_O|\go 2(\hbar c/E) |d\phi_B/ds|$ 
(where $n_X,~n_O$ are the indices of refraction of the two modes); 
this condition breaks down at $r=r_{\rm pl}$, which is much greater than
the stellar radius for all parameter regimes of interest in this paper.
The observed Stokes parameters (normalized to the total intensity $I$)
are given by $Q/I=(2P_{\rm con}-1)p_e\cos 2\phi_B(r_{\rm pl})$
and $U/I=(2P_{\rm con}-1)p_e\sin 2\phi_B(r_{\rm pl})$,
where $p_e=(I_X^\zero-I_O^\zero)/(I_X^\zero+I_O^\zero)$ is the 
``intrinsic'' polarization fraction at emission, and $\phi_B(r_{\rm pl})$
is the azimuthal angle of the magnetic field at $r=r_{\rm pl}$ 
[For $r_{\rm pl}\ll c/\Omega$, or for spin frequency $\ll 70$~Hz,
$\phi_B(r_{\rm pl})\simeq \pi+\phi_B(R)$]. We calculate the observed 
spectral fluxes $F_I,~F_Q,~F_U$ (associated with the intensities $I,~Q,~U$)
using the standard procedure, including the effect of general 
relativity \cite{pechenicketal83}.

Figure~\ref{fig:f3} 
shows the total flux and polarization ``light curves''
generated from the hot polar cap of a rotating NS.
As the star rotates, the plane of linear polarization rotates. 
The light curves obviously depend on the geometry (specified by the angles
$\beta$ and $\gamma$). Note that $F_Q$ for low-energy ($E\lo 1$~keV)
photons is opposite to that for high-energy photons ($E\go 4$~keV), which
implies that the planes of linear polarization at low and high energies are
orthogonal. This feature also manifests in the phase-averaged linear
polarization (see Fig.~\ref{fig:f4}). This is an unique signature of photon
mode conversion induced by vacuum polarization.


\nobreak
\begin{figure}
\includegraphics[height=8cm]{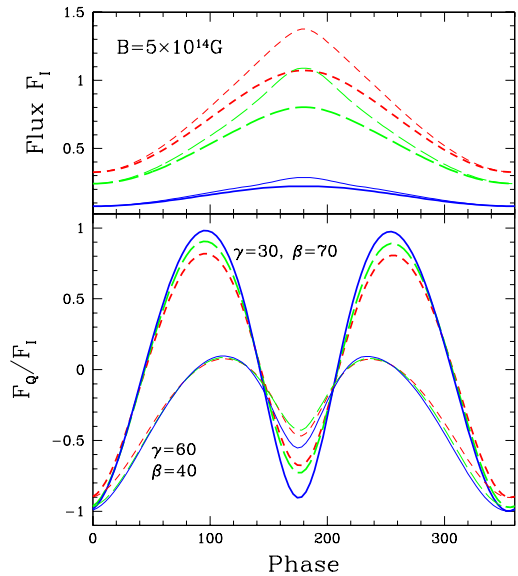}
\caption{
Same as Fig.~\ref{fig:f3}, except for
$B=5\times 10^{14}$~G and $T_{\rm eff}=5\times 10^6$~K.
The different curves are for different photon energies (1~keV dashed,
2~keV long-dashed, 5~keV solid lines). The thicker 
curves are for $\gamma=30^\circ$,
$\beta=70^\circ$, and the thin curves for $\gamma=60^\circ$, 
$\beta=40^\circ$. In contrast to Fig.~\ref{fig:f3}-\ref{fig:f4},
in the magnetar field regime, the planes of linear polarization at different
photon energies coincide with each other.
\label{fig:f5}}\end{figure}

In the magnetar field regime, $B>B_l$, the vacuum resonance lies between
the photospheres of the two modes.
At low energies (e.g., $E\lo 1$~keV), no mode
conversion occurs at the vacuum resonance, and vacuum polarization makes
the X-mode photosphere lie above (i.e. at lower density) 
its ``original'' location (i.e. when
vacuum polarization is turned off) because the photon opacity
exhibits a large spike at the vacuum resonance \cite{laiho02}. At high
energies (e.g., $E\go 4$~keV), the effective X-mode photosphere lies very 
near the vacuum resonance. For both low and high energies, the emergent
radiation is dominated by the X-mode (for $\theta_{kB}$ not too close to 0). 
Therefore the planes of linear polarization at different energies coincide and
evolve ``in phase'' as the star rotates (see Fig.~\ref{fig:f5}).
These polarization signals are qualitatively different 
from the ``normal'' field regime. Note that in the
magnetar field regime, vacuum polarization significantly affects the total spectral flux from the atmosphere \cite{laiho02,holai03,laiho03}:  
(i) Vacuum polarization makes the effective decoupling density of X-mode
photons (which carry the bulk of the thermal energy) smaller, thereby depleting
the high-energy tail of the spectrum and making the spectrum closer to
black-body; (ii) vacuum polarization suppresses the proton cyclotron line
and other spectral line features \cite{hoetal03} in the spectra by making 
the decoupling depths inside and outside the line similar. 
As we discussed in previous papers \cite{holai03,hoetal03}, 
the absence of lines in the observed thermal spectra of several magnetar 
candidates \cite{juettetal02} may be considered as a proof of the vacuum
polarization effect at work in these systems. Measurement of X-ray polarization
would provide an independent probe of the magnetic fields of these objects.

Finally, we note that although the specific results shown in
Figs.~\ref{fig:f3}-\ref{fig:f5}
refer to emission from a hot polar cap on the NS,
we expect the vacuum polarization signature (e.g., that the planes of 
linear polarization at $E\lo 1$~keV and at $E\go 4$~keV are orthogonal
for $B\lo 7\times 10^{13}$~G) to be present in more complicated models
(e.g. when several hot spots or the whole stellar surface contribute to the
X-ray emission). This is because the polarization-limiting radius (due to
vacuum polarization in the magnetosphere) lies far away from the star 
(see above), where rays originating from different patches of the star 
experience the same dipole field \cite{heyletal03}. Our results
therefore demonstrate the unique potential of X-ray
polarimetry\cite{costaetal01} in 
probing the physics under extreme conditions (strong gravity and magnetic
fields) and the nature of various forms of NSs.

This work was supported by NASA grant NAG 5-12034
and NSF grant AST 0307252.


\end{document}